\documentclass[a4paper,11pt]{article}
\usepackage{pos}

\newcommand{\bibi}[8]{\bibitem{#1} #2, \emph{#3}, #4, \href{https://doi.org/#5}{\emph{#6} \textbf{#7} #8}}
\newcommand{\bibinodoi}[8]{\bibitem{#1} #2, \emph{#3}, #4, \href{#5}{\emph{#6} \textbf{#7} #8}}

\title{The ablation of gas clouds by blazar jets and the long-lasting flare in CTA 102}
 \ShortTitle{Ablation of gas clouds by blazar jets}

\author*[a,b]{Michael Zacharias}
\author[c]{Jonathan Heil}
\author[b]{Markus B\"ottcher}
\author[d]{Felix Jankowsky}
\author[e]{Jean-Philippe Lenain}
\author[d]{Stefan Wagner}
\author[f]{Alicja Wierzcholska}

\affiliation[a]{Laboratoire Univers et Théories, Observatoire de Paris, Université PSL, CNRS, Université de Paris, \\ 5 pl Jules Janssen, 92190 Meudon, France}

\affiliation[b]{Centre for Space Science, North-West University, \\ Potchefstroom, 2520, South Africa}

\affiliation[c]{Ruhr Astroparticle and Plasma Physics Center (RAPP Center), Insitut f\"ur Theoretische Physik IV, Ruhr-Universit\"at Bochum, \\ D-44780 Bochum, Germany}

\affiliation[d]{Landessternwarte, Universit\"at Heidelberg, \\ K\"onigstuhl 13, D-69117 Heidelberg, Germany }

\affiliation[e]{Sorbonne Universit\'e, Universit\'e Paris Diderot, Sorbonne Paris Cit\'e, CNRS/IN2P3, Laboratoire de Physique Nucl\'eaire et de Hautes Energies, LPNHE, \\ 4 Place Jussieu, F-75252 Paris, France}

\affiliation[f]{Institute of Nuclear Physics, Polish Academy of Sciences, \\ PL-31342 Krakow, Poland}

\emailAdd{michael.zacharias@obspm.fr}
\emailAdd{mzacharias.phys@gmail.com}

\abstract{Long-lasting, very bright multiwavelength flares of blazar jets are a curious phenomenon. The interaction of a large gas cloud with the jet of a blazar may serve as a reservoir of particles entrained by the jet. The size and density structure of the cloud then determine the duration and strength of the particle injection into the jet and the subsequent radiative outburst of the blazar. In this presentation, a comprehensive parameter study is provided showing the rich possibilities that this model offers. Additionally, we use this model to explain the 4-months long, symmetrical flare of the flat spectrum radio quasar CTA 102 in late 2016. During this flare, CTA 102 became one of the brightest blazars in the sky despite its large redshift of $z=1.032$.
}

\FullConference{37$^{\rm{th}}$ International Cosmic Ray Conference (ICRC 2021)\\
		July 12th -- 23rd, 2021\\
		Online -- Berlin, Germany}

\begin{document}
\maketitle

\section{Introduction}
The spectral energy distribution of blazars, the relativistically beamed version of radio-loud active galactic nuclei (AGNi) \cite{br74}, is characterized by the famous double-hump structure. The first component peaks between the IR and X-ray energies and is attributed to synchrotron emission of relativistic electrons (as well as positrons). The second component peaks in the MeV to TeV $\gamma$-rays. Within a leptonic scenario, this peak is attributed to inverse-Compton emission of the same electron population scattering ambient photons, such as their own synchrotron emission (synchrotron-self Compton, SSC), or thermal photons from the accretion disk, the broad-line region (BLR) or the dusty torus (DT). In hadronic scenarios, the second component may be due to synchrotron emission of relativistic protons, or synchrotron emission of pairs created through the pion- or Bethe-Heitler-induced cascade. For a recent review on the emission processes in blazars, see \cite{c20}.

Blazars exhibit variability on various time scales and with varying intensity \cite{z18}. Typically, one assumes a compact region somewhere in the jet, where particles get accelerated and radiate. A change in the parameters of this compact region -- for example, varying particle densities, magnetic field, etc -- results in the outburst \cite{db14}. However, the nature of these parameter variations is not necessarily specified. A natural source of both particle and magnetic field variations would be the accretion disk. However, the connection between the disk and the jet is not yet fully understood and remains a topic of active research, e.g. \cite{vf21}. Another source of additional particles is the interaction of the jet with an obstacle -- a star or a cloud -- that moves in the way of the jet, e.g. \cite{bea12,tabr19,pbr19}. If the obstacle is fast and compact, such as a star close to the base of the jet, rapid variability may occur \cite{bea12}. On the other hand, if the object is larger in size, such as the cloud-like objects one can find in star-forming regions, the interaction between the jet and the cloud may take a long time. In turn, the ablation of the cloud and the subsequent injection of the cloud particles into the jet proceeds on these longer time-scales and depends strongly on the density structure of the cloud. This may result in a long-lasting flare (order of months or more).

Below, we will summarize the ablation model, for which details can be found in \cite{hz20}. Furthermore, we provide the application of the model to the giant flare in CTA~102 \cite{zea17,zea19}, naturally describing the long-lasting flaring state.

\section{The cloud ablation model}
\begin{figure}[th]
\centering 
\includegraphics[width=0.35\textwidth]{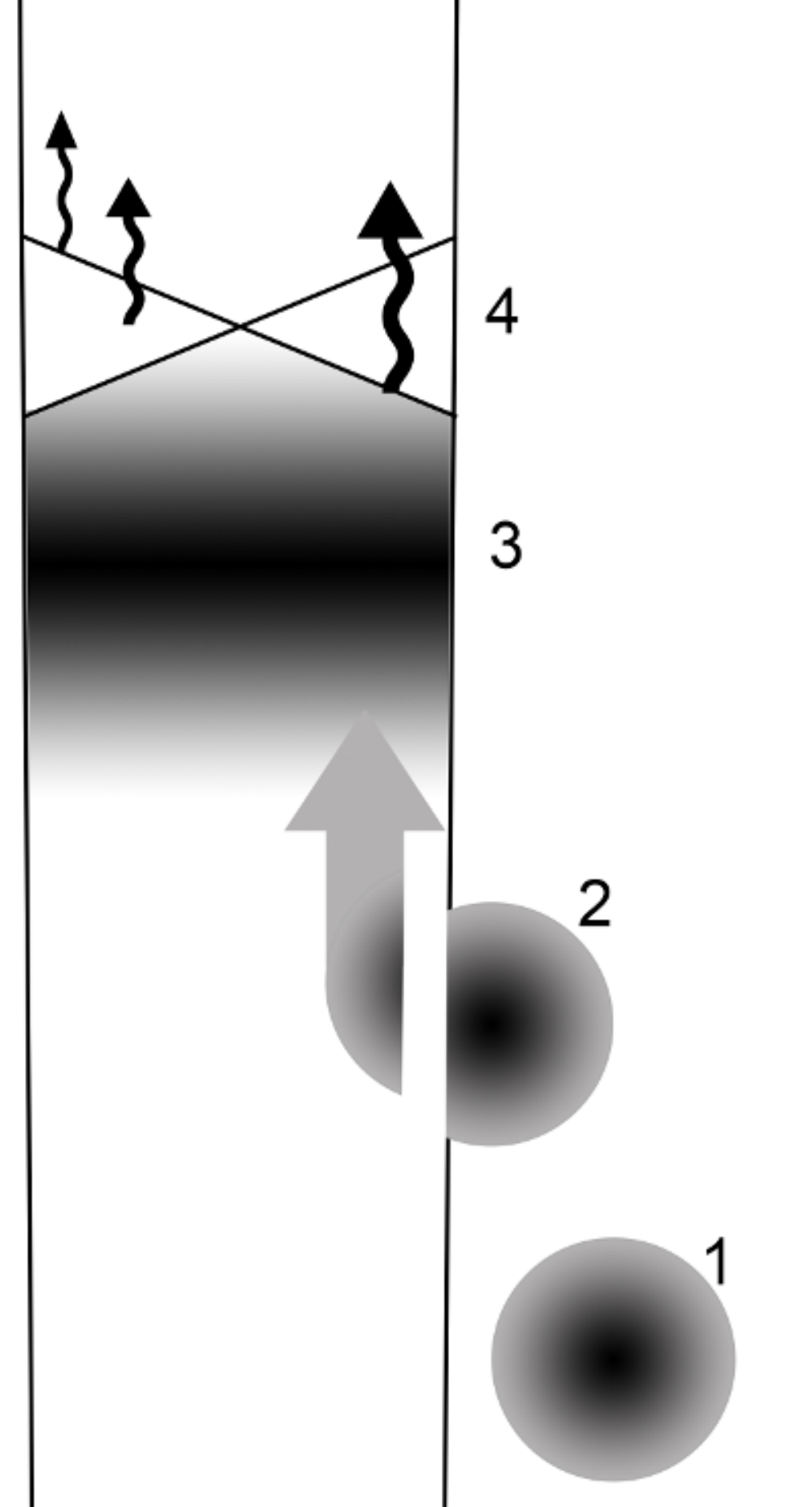}
\caption{Sketch of the ablation process (not to scale). (1) The cloud approaches the jet and (2) is ablated while entering the jet. (3) The cloud material is mixed into the jet flow providing a specific density enhancement. (4) At a downstream shock, the particles are accelerated and radiate. Figure from \cite{hz20}.
}
\label{fig:sketch}
\end{figure} 
In case of hydrostatic equilibrium, that is the cloud's density structure is governed by the interplay of the self-gravitational pull and the thermal pressure, the particle density $n$ as a function of cloud radius $r$ can be approximated with a simple form:

\begin{align}
	n(r) = \frac{n_0}{1+(r/r_0)^2}
	\label{eq:clouddens}.
\end{align}
It depends on the central density $n_0$, and the scale height $r_0$, which depends on $n_0$ and the cloud's temperature $T_c$ according to

\begin{align}
	r_0 = 4\times 10^{12}\,\left( \frac{T_c}{140\,\mbox{K}} \right)^{1/2} \left( \frac{n_0}{10^{15}\mbox{cm}^{-3}} \right)^{-1/2}\,\mbox{cm}
	\label{eq:scaleheight}.
\end{align}
As a cloud cannot be infinitely large, we define an outer radius $R_c>r_0$ beyond which the cloud density is zero.

The ablation process commences, if the jet's ram pressure is high enough to induce a shock in the entering cloud material, which is also fast enough to cross the cloud before the latter has exited the jet on the other side. This gives two conditions:

\begin{align}
	\Gamma &\gtrsim 11 \left( \frac{T_c}{140\,\mbox{K}} \right)^{1/2} \left( \frac{n_0}{10^{15}\mbox{cm}^{-3}} \right)^{1/2} \left( \frac{a}{0.1} \right)^{-1/2} \left( \frac{n_j}{10^3\mbox{cm}^{-3}} \right)^{-1/2} \label{eq:limGam} \\
	v_s &> 3.2\times 10^{3} \left( \frac{r_0}{4\times 10^{12}\mbox{cm}} \right)  \left( \frac{v_c}{2\times 10^7\mbox{cm/s}} \right)  \left( \frac{R_j}{2.5\times 10^{16}\mbox{cm}} \right)^{-1}\ \mbox{cm/s} \label{eq:limshock},
\end{align}
where $\Gamma$ is the jet's bulk Lorentz factor, $a$ is the proton-to-electron ratio in the jet, $n_j$ is the jet density, $v_s$ is the shock speed, $v_c$ is the cloud speed, and $R_j$ is the jet radius. Note the misprints in \cite{hz20}, their Eqs.~(8) and (9), where the ``smaller'' signs should actually be ``larger'' signs. 

\begin{figure}[th]
\begin{minipage}{0.49\linewidth}
\centering \resizebox{\hsize}{!}
{\includegraphics{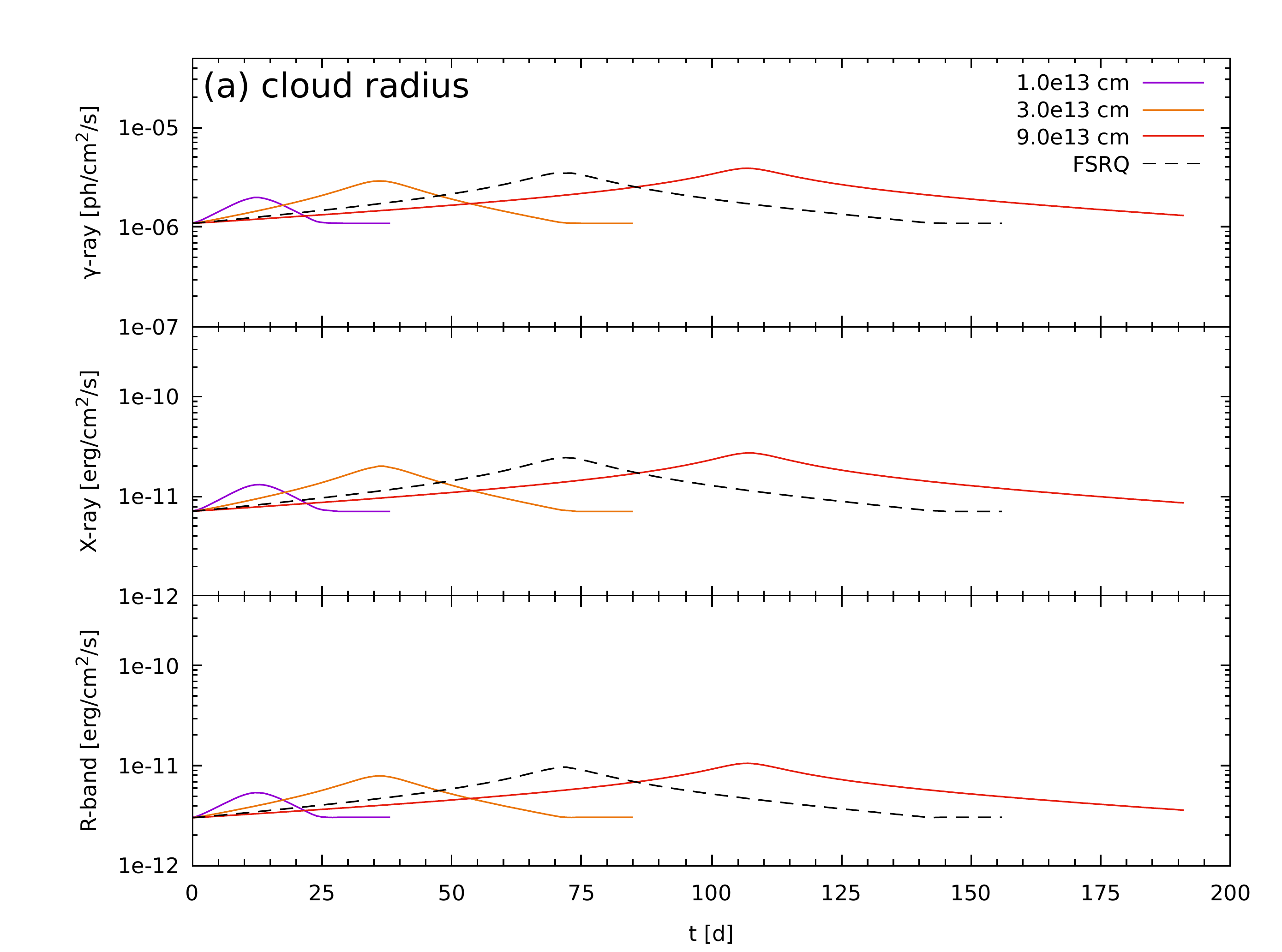}}
\end{minipage}
\hspace{\fill}
\begin{minipage}{0.49\linewidth}
\centering \resizebox{\hsize}{!}
{\includegraphics{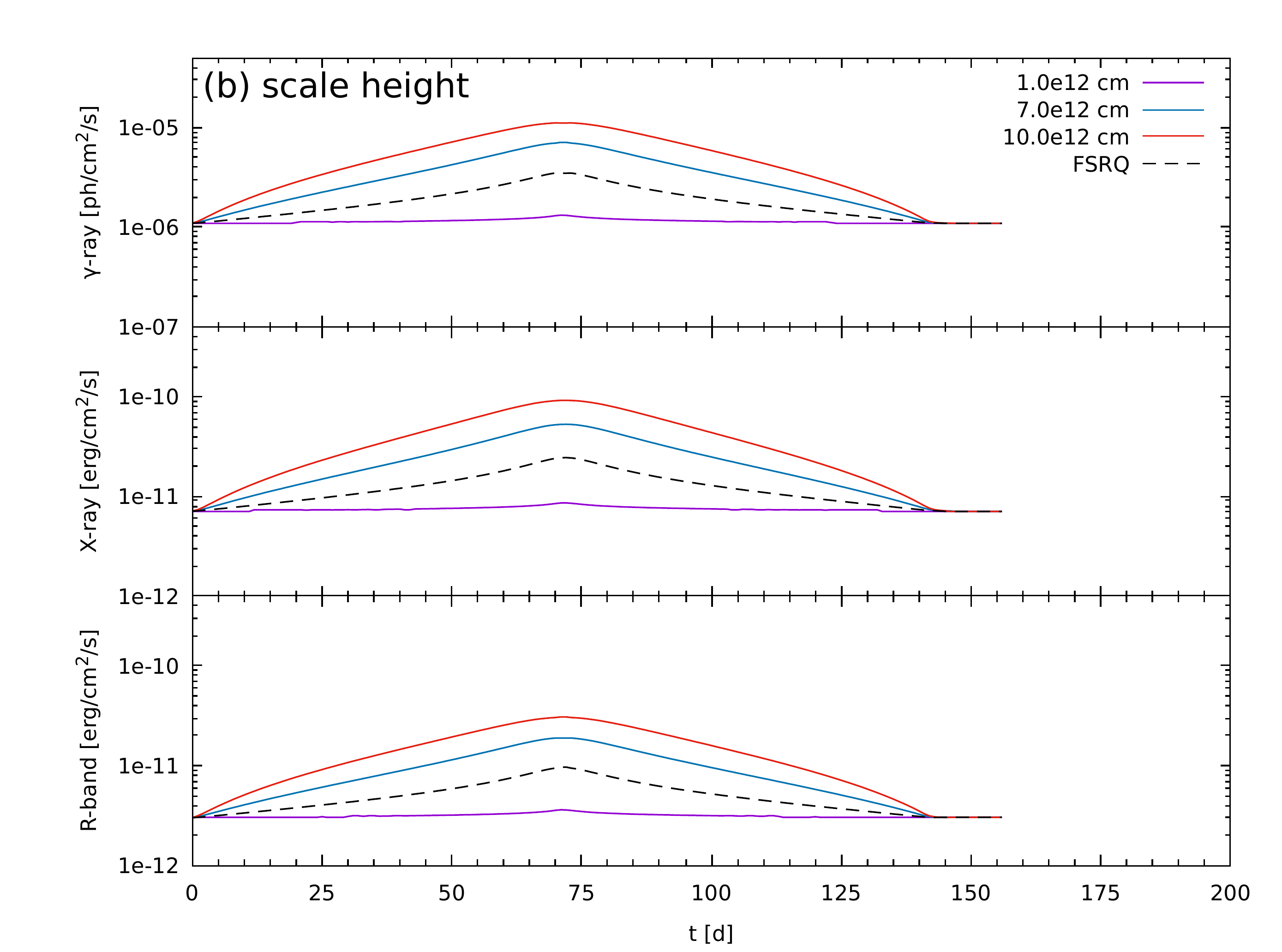}}
\end{minipage}
\newline
\begin{minipage}{0.49\linewidth}
\centering \resizebox{\hsize}{!}
{\includegraphics{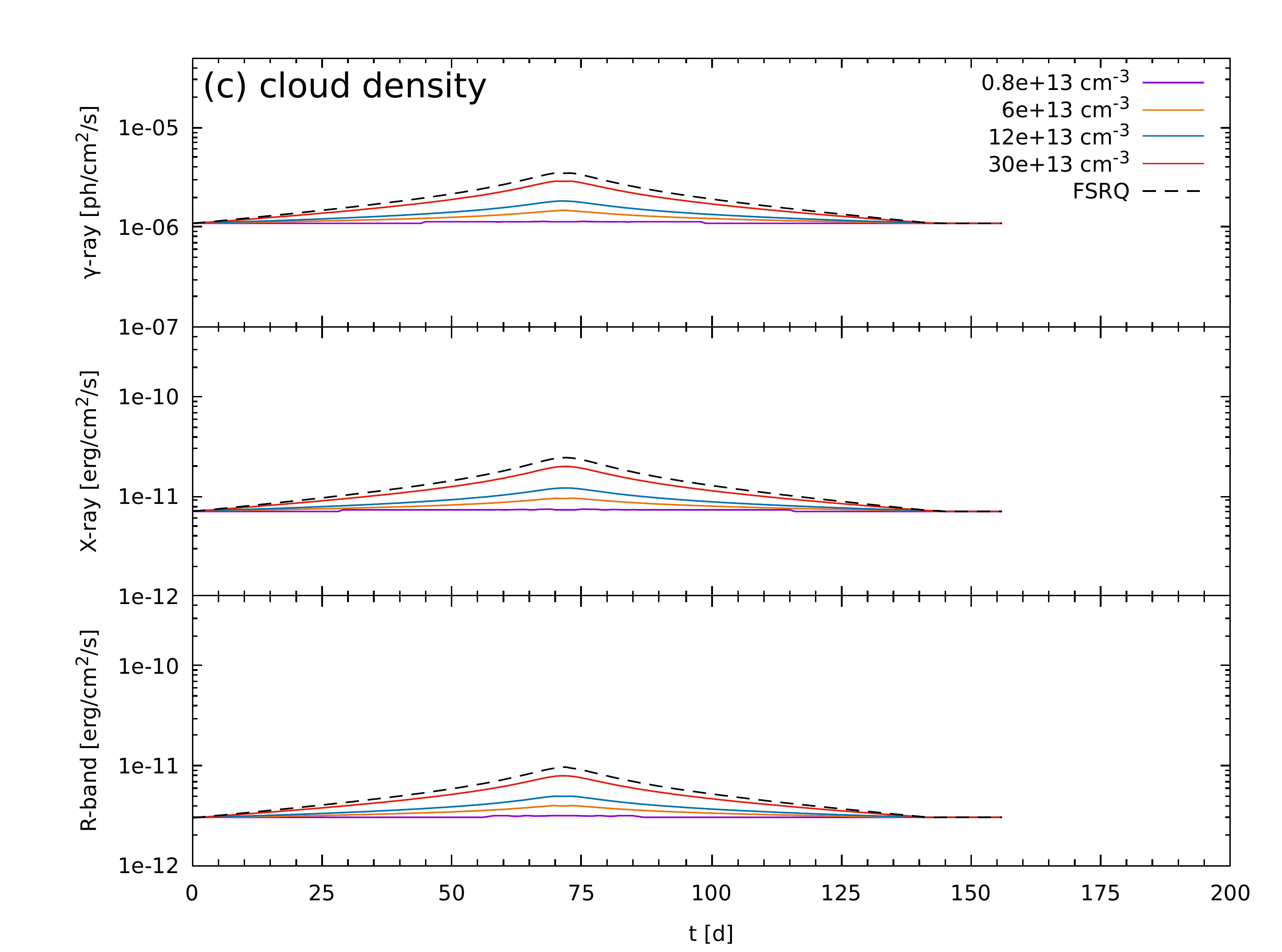}}
\end{minipage}
\hspace{\fill}
\begin{minipage}{0.49\linewidth}
\centering \resizebox{\hsize}{!}
{\includegraphics{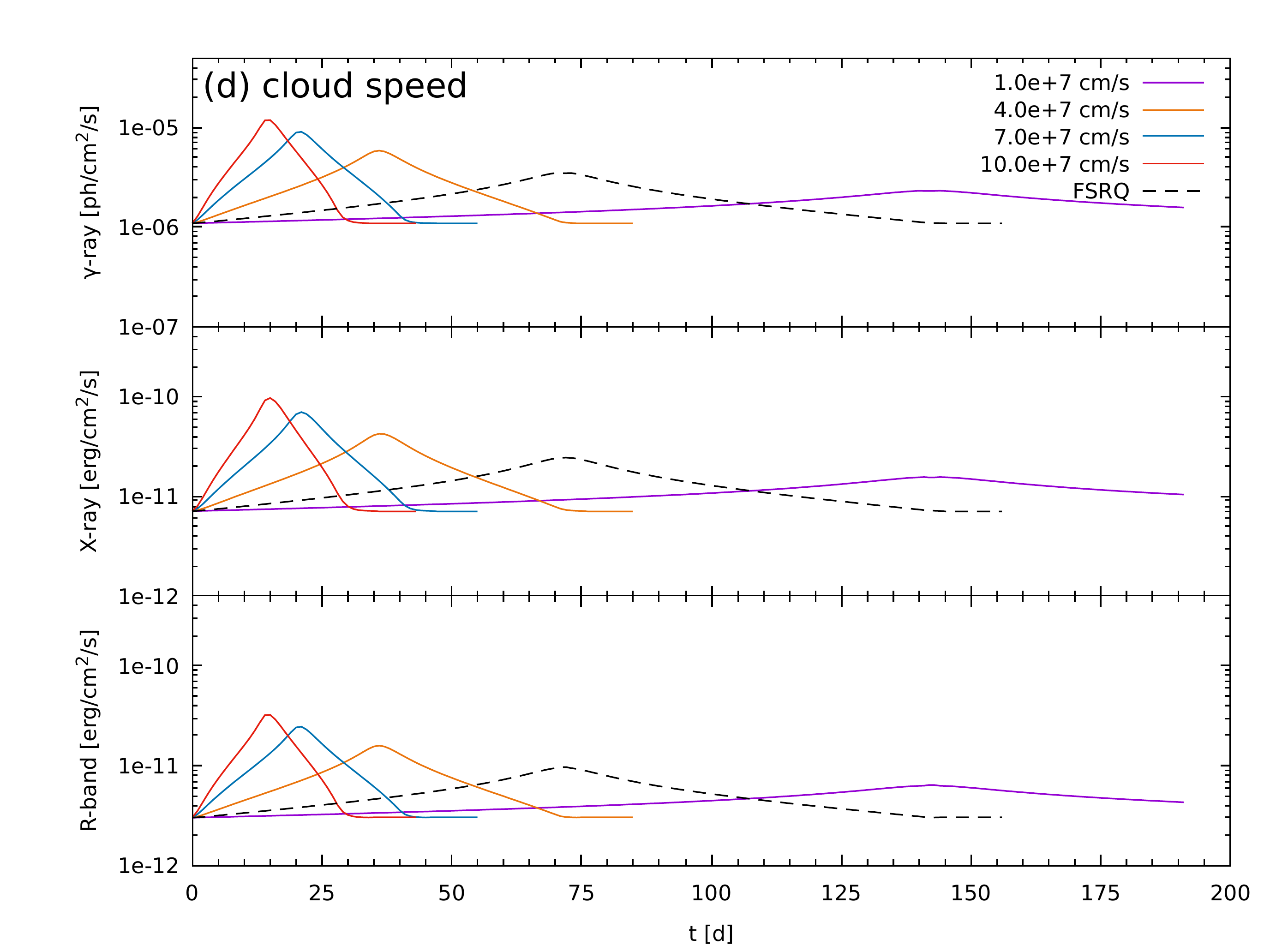}}
\end{minipage}
\caption{Lightcurves in the observer's frame for different cloud parameters. In each panel, lightcurves in the $\gamma$-ray, X-ray and R band are shown for: (a) cloud radius $R_c$, (b) scale height $r_0$, (c) cloud density $n_0$, and (d) cloud speed $v_c$. The dashed black lightcurve is the same baseline simulation in each panel. 
Note the logarithmic y-axes. Figure from \cite{hz20}.
}
\label{fig:parastud1}
\end{figure} 
A sketch of the principle steps in the ablation process is shown in Fig.~\ref{fig:sketch}. After approaching the jet, the cloud is ablated, and the cloud matter is mixed into the jet material. The rate of cloud particles entering the jet is given in the comoving (primed) frame of the jet by

\begin{align}
	\frac{\mbox{d}N(t^{\prime})}{\mbox{d}t^{\prime}} = \Gamma \pi n_0v_cr_0^2 \ln{\left\{ \frac{(\Gamma t_0)^2+(\Gamma t_R)^2}{(\Gamma t_0)^2 + (\Gamma t_R - t^{\prime})^2} \right\}}
	\label{eq:injrate},
\end{align}
where $t_0 = r_0/v_c$ and $t_R = R_c/v_c$. These particles are accelerated along with the jet material at a downstream shock, where the main radiative emission takes place. For the sake of a simple model, we do not consider the shocks involved in the ablation processes as acceleration and emission sites for the material, as their evolution is complex and requires detailed plasma physical simulations, which are beyond the scope of this study.

The theoretically expected light curves are shown in Fig.~\ref{fig:parastud1}. We show for the $\gamma$-ray, the X-ray and the optical R-band the light curves for different parameter settings. The cloud radius $R_c$ influences mainly the duration of the event, while the scale height has a significant influence on the light curve shape. This is easy to understand, as in this case the outer radius is constant, and if $r_0$ approaches $R_c$, the density structure becomes more and more uniform explaining the different light curve shapes. Additionally, in this case a higher number of particles is enclosed within $R_c$ compared to cases with smaller $r_0$ explaining the difference in the normalization. The central density $n_0$ obviously influences the normalization of the light curve, while the cloud speed changes the duration of the event and the light curve normalization. As a slower cloud takes longer to penetrate the jet, less particles are injected per time step explaining the reduced normalization.

\section{Application to CTA~102}
\begin{figure}[th]
\centering 
\includegraphics[width=0.65\textwidth]{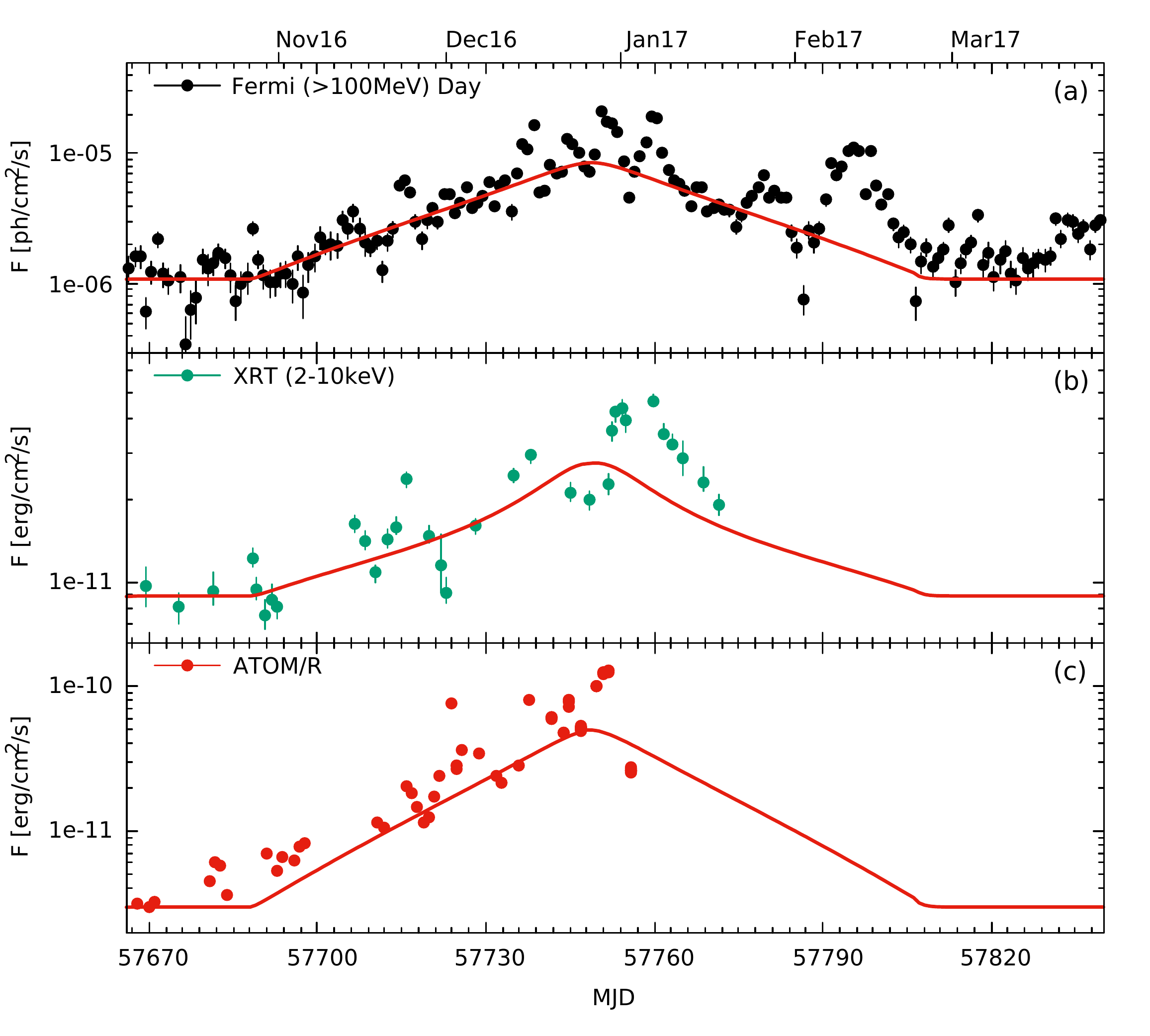}
\caption{Data from CTA~102 in the $\gamma$-ray band from Fermi-\textit{LAT} (top), in the X-ray band from Swift-XRT (middle), and in the optical R-band from ATOM (bottom). The red lines mark the model light curve.
Note the logarithmic y-axis. Figure from \cite{zea17}.
}
\label{fig:cta102}
\end{figure} 
CTA~102 is a flat spectrum radio quasar at a redshift of $1.032$. Between early 2016 and early 2018 it underwent several major outbursts, of which the brightest was roughly in the middle of this time span in early January 2017. It became so bright that it was briefly the brightest source in $\gamma$ rays, and became visible in the optical band to the human eye through small telescopes. The multiwavelength light curves of the main event are shown in Fig.~\ref{fig:cta102}.

We use the cloud ablation model to reproduce the long-term trend, and achieve an excellent fit. The shorter-term spikes are not of our concern, and could be caused by sub-structures in the cloud, or instabilities in the jet during the ablation process. From the fit to the data, we can deduce the cloud parameters \cite{zea17,zea19}. We obtain for $v_c=5.12\times 10^8$cm/s (which is the orbital speed around the black hole at the boundary of the BLR), a cloud radius of $1.3\times 10^{15}$cm, a density of $n_0 = 2.54\times 10^{8}$cm$^{-3}$, and a scale height of $r_0=1.6\times 10^{14}$cm. In turn, the temperature is $T_c = 0.5$K.

This temperature is clearly too low for a cloud in close orbit around an active galactic nucleus. However, in this determination we have assumed that all cloud particles enter the jet and also take part in the emission processes. This is not realistic. Firstly, a shock only accelerates $10\%$ or less of the particles that cross it. Secondly, the ablation process itself will not be as smooth as described above. In fact, it is quite likely that the shocks forming during the ablation process eject a substantial amount of particles away from the interaction -- and away from the jet. This is however even less quantifiable than the number of shock accelerated particles. However, we can confidently say that the actual density of the cloud must have been higher than what we deduce from the fit to the light curves, and in turn the temperature must increase as well following Eq.~(\ref{eq:scaleheight}).

The cloud parameters listed above can be compared to known cloud types. They correspond well to clouds in star-forming regions, and to a lesser degree to BLR clouds and astrospheres of giant stars. While blazars are typically hosted in elliptical galaxies, which are rather devoid of gas and dust, such clouds could still be present. Firstly, the active state of the nucleus proves the presence of gas in the central region. Secondly, elliptical galaxies probably form from the merger of spiral galaxies, and their gas and dust is partially funneled into the core of the new galaxy, where it could interact with the jet.

\section{Summary}
The ablation of gas clouds by the relativistic jet of an active galaxy is a viable process to produce long-lasting flares. We have summarized the work of \cite{hz20}, where the details of the model are provided. If a gas cloud approaches the jet, the interaction will ablate the cloud material and carry it along in the jet flow. Given the density structure in the cloud, this provides a unique density profile in the jet. At a shock downstream from the interaction site, the jet and cloud matter is accelerated and radiates.

We have shown the resulting light curve profiles for different cloud parameters (for a more comprehensive study, see \cite{hz20}). It turns out that the ablation model is capable of explaining different flare profiles in terms of flare duration, intensity, and shape. We note again that the model aims to explain long-lasting flares of weeks to months in duration. 

The model has been applied to the long-lasting, and surprisingly symmetric flare in CTA~102 from late 2016 to early 2017. The reproduction of the multiwavelength data is excellent \cite{zea17,zea19}. The inferred cloud parameters are similar to those of clouds in star forming regions, which is a reasonable scenario in the centre of an active galaxy.


%
%
%

\end{document}